\documentclass{aa}  

\def\sgr{Sgr A$^{\star}$\xspace}

\def\flaremodel{\textsc{flaremodel}\xspace}
\def\lmfit{\textsc{lmfit}\xspace}
\def\tinj{\textrm{inj}}
\def\tesc{\textrm{esc}}
\def\tsyn{\textrm{syn}}
\def\tssc{\textrm{SSC}}
\def\tic{\textrm{IC}}
\def\tin{\textrm{in}}
\def\tS{\textrm{S}}

\usepackage{graphicx}
\usepackage{txfonts}
\usepackage{color}

\begin{document}

   \title{Flaremodel: An open-source Python package for one-zone numerical modelling of synchrotron sources}

    \author{
Y. ~Dallilar
\inst{1}
\and
S.~von Fellenberg\inst{1}
\and 
M.~Bauboeck\inst{1}
\and
P.T.~de Zeeuw\inst{2,1}
\and 
A.~Drescher\inst{1, 3}
\and
F.~Eisenhauer\inst{1}
\and 
R.~Genzel\inst{1, 4}
\and
S.~Gillessen\inst{1}
\and
M.~Habibi\inst{1}
\and 
T.~Ott\inst{1}
\and 
G.~Ponti\inst{5, 1}
\and
J.~Stadler\inst{1}
\and 
O.~Straub\inst{1}
\and 
F.~Widmann\inst{1}
\and
G.~Witzel\inst{6}
\and
A.~Young\inst{1}
}

   \institute{
%1  
Max Planck Institute for extraterrestrial Physics,
Giessenbachstrasse~1, 85748 Garching, Germany
%2
\and Sterrewacht Leiden, Leiden University, Postbus 9513, 2300 RA
Leiden, The Netherlands
%3
\and Department of Physics, Technical University Munich, James-Franck-Strasse 1,  85748 Garching, Germany
%4
\and Departments of Physics and Astronomy, Le Conte Hall, University
of California, Berkeley, CA 94720, USA
%5
\and INAF-Osservatorio Astronomico di Brera, Via E. Bianchi 46, I-23807 Merate (LC), Italy
%6
\and Max Planck Institute for Radio Astronomy, Auf dem H\"ugel 69, 53121 Bonn, Germany
}
   \date{Received October 15, 2021; accepted November 17, 2021}

% \abstract{}{}{}{}{} 
% 5 {} token are mandatory
 
  \abstract
    { Synchrotron processes, the radiative processes associated with the interaction of energetic charged particles with magnetic field, are of interest in many areas in astronomy, from the interstellar medium to extreme environments near compact objects. Consequently, observations of synchrotron sources carry information on the physical properties of the sources themselves and those of their close vicinity. In recent years, novel observations of such sources with multi-wavelength collaborations reveal complex features and peculiarities, especially near black holes. Exploring the nature of these sources in more detail necessitates numerical tools complementary to analytical one-zone modelling efforts. In this paper, we introduce an open-source Python package tailored to this purpose, \flaremodel. The core of the code consists of low-level utility functions to describe physical processes relevant to synchrotron sources, which are written in C for performance and parallelised with OpenMP for scalability. The Python interface provides access to these functions and built-in source models are provided as a guidance. At the same time, the modular design of the code and the generic nature of these functions enable users to build a variety of source models applicable to many astrophysical synchrotron sources. We describe our methodology and the structure of our code along with  selected examples demonstrating capabilities and options for future modelling efforts.
}

   \keywords{Radiative transfer; radiation mechanisms: non-thermal; stars: black holes}

   \maketitle
%
%-------------------------------------------------------------------

\section{Introduction}

Synchrotron processes describe the energy loss of charged particles in the presence of a magnetic field. Although there are many applications of synchrotron processes in astrophysics, the most luminous synchrotron sources originate near extreme environments, such as black holes, and often in the form of jets. Measurement of the spectral energy distribution (SED) and its modelling allow  extraction of the physical conditions in the vicinity of such sources.

A prime example is \sgr, the nearest supermassive black hole located at a distance of 8.27 kpc towards the centre of our Galaxy \citep{GravityCol19, Do2019, GravityCol21} with a mass of $4.3\times10^6 \textrm{ M}_\odot$. Recent interferometric observations with GRAVITY pinpoint the origin of \sgr flares to localised `hot-spots' orbiting at several $R_{\tS}$ ($=1.3\times10^{12}$ cm) in the accretion flow of the black hole, an interpretation also supported by polarisation signatures \citep{GravityCollaboration2018_orbital, GravityCollaboration2020_orbital, GravityCollaboration2020_polariflares}. However, possible emission mechanisms for \sgr, or accreting black holes in general, are still subject to debate, particularly from near-infrared (NIR) to X-ray. Proposed scenarios include a mixture of synchrotron, synchrotron self-Compton (SSC), and inverse Compton (IC) models \citep{Dodds-Eden09,eckart12, dibi14,2021arXiv210701096A}. A variety of temporal models for the \sgr flares have also been developed with various approaches; see for example \cite{YusefZadeh08}, \cite{Dodds-Eden10}, and \cite{Witzel21}.  

Interpreting observations from state-of-the-art observational facilities necessitates flexible numerical utilities that can accommodate easy development of various source models to disentangle these emission scenarios and to explore the nature of these sources in detail. The Python package \flaremodel described in this paper is designed for this purpose\footnote{Available at \url{https://github.com/ydallilar/flaremodel}}. The code is developed around generic low-level utility functions which are written in C for performance. These functions constitute the core of \flaremodel. For convenience and accessibility, we developed the user interface in Python. The user interface includes exported utility functions and readily available source models as a guidance, some of which are described in this paper. Our code has also been used in \cite{2021arXiv210701096A} to investigate emission mechanisms of a peculiar NIR/X-ray flare of \sgr. 

Thanks to the modular design of \flaremodel, it is also possible to build a variety of models applicable to astrophysical synchrotron sources with the help of these utility functions. This feature allows users to explore a wide selection of models in a short time and is one of the main strengths of our code in comparison to similar alternatives, such as \textsc{agnpy} \citep{cosimo_nigro_2020_4311271}. Options for spatial and temporal modelling are also available to a limited extent as described later in this paper.

Another distinguishing feature of this code is the support for parallel computations. In terms of overheads, OpenMP implementation in C is far more efficient than alternative implementations available in Python for parallel computations. Therefore, multi-threading is implemented at a lower level in C functions using OpenMP. There may be situations where higher level multi-threading is desired on the Python side but the efficiency is inherently hindered by the Python global interpreter lock (GIL). As the C functions in our code bypass GIL, this provides flexibility for developing source models targeting high-end computing hardware. As an experimental feature, we provide a Python interface to extend the functionality with algorithms targeting graphics processing units (GPUs) with OpenCL implementation. Currently, there is only built-in support for IC model calculations.

We further provide an SED fitting module built on top of the Python fitting package \lmfit \citep{newville_matthew_2014_11813}. This package provides a user-friendly interface to a range of optimisation methods available within the Python ecosystem ranging from least-square minimisation to Markov-Chain Monte Carlo methods. As \flaremodel opens many possibilities towards source modelling, we do not attempt to cover all the bases for fitting. Users can extend the fitting class following the examples in our documentation. Alternatively, users may opt for other fitting packages of their preference for which we do not provide explicit support and examples are not included in our documentation.

The organisation of this paper is as follows. In Sect.~\ref{sec:method}, we describe the structure of our code and physical processes implemented therein. Section~\ref{sec:test} includes selected test problems and discusses selected cases for which we demonstrate this code.

\section{Methodology} \label{sec:method}

\subsection{Electron distributions} \label{sec:edist}

Physical processes and interactions implemented in our code are limited to electrons. We provide a selection of built-in electron distributions with this code; thermal, kappa, and a family of power-law distributions. Power-law distributions are either in the form of a plain power law, a broken power law, or their exponential cutoff variants. The equations of the built-in electron distributions and their normalisation are detailed in Appendix~\ref{sec:ap_edist}.

Our code includes a modular low-level C interface which is used to define built-in electron distributions. As utility functions in our code do not contain instructions specific to the electron distributions, the same interface can be used to add new electron distributions and these will be readily available for source models. Our code documentation covers the guidelines for defining new electron distributions in detail.

\subsection{Synchrotron emissivity and absorption coefficients} \label{sec:j_a_nu}

We calculate the synchrotron emissivity and absorption coefficients ($j_{\nu}$, $\alpha_{\nu}$) per steradian as follows,
\begin{align}
    j_{\nu} &= \frac{1}{4\pi} \int^{\gamma_{\max}}_{\gamma_{\min}} P_{e} n_e(\gamma) d\gamma, \\
    \alpha_{\nu} &= \frac{1}{8\pi m_e \nu^2} \int^{\gamma_{\max}}_{\gamma_{\min}} P_e \left(\frac{\textrm{d}  n_e(\gamma)}{\textrm{d} \gamma}-\frac{2 n_e(\gamma)}{\gamma}\right) \textrm{d}\gamma, \label{eq:abs_coef}
\end{align}
where $m_e$ is the mass of an electron, $n_e(\gamma)$ is the electron density, $\gamma$ is the Lorentz factor, and $\nu$ is the frequency of the emitted/absorbed radiation. The power emitted by a single electron ($P_e$) is given by
\begin{equation}
    P_e = \frac{\sqrt{3} e^3 B \sin\phi}{m_e c^2} F(X).
\end{equation}
Here, $e$ is the charge of an electron, $c$ is the speed of light, $B$ is the magnetic field, and $\phi$ is the pitch angle. The $F(X)$  function is defined as 
\begin{equation}
    F(X) = X \int^{\infty}_X K_{5/3}(x) dx,
\end{equation}
where $X=\nu/\nu_c$ and $K(x)_{5/3}$ is the modified Bessel function of order 5/3. The critical frequency ($\nu_c$) is 
\begin{equation} \label{eq:nu_crit}
    \nu_c = \frac{3}{4\pi} \frac{eB\gamma^2 \sin\phi}{m_e c}.
\end{equation}

Evaluating the $F(X)$ function is numerically expensive. Instead, we use pre-evaluated values of the function and perform interpolation at run-time in the range $10^{-4}<X<10$. Beyond these limits, we use the approximations from \cite{RL79}. The reader may refer to \cite{BlumenthalGould70} for the detailed formulation of synchrotron radiation.

There are two different methods implemented in our code; \textit{brute} or \textit{userdist}. The only difference between these two methods is the estimate of electron distribution derivative as required by Eq.~(\ref{eq:abs_coef}). While the \textit{brute} method uses the analytical form of a given electron distribution, the \textit{userdist} method estimates the quantity from a user-defined numerical grid. Therefore, the \textit{brute} method is more accurate and is the default method. The \textit{userdist} method is originally motivated by non-trivial electron distributions arising from the temporal evolution of electrons, which we describe in Sect.~\ref{sec:temp}.

Our default integration scheme relies on an equidistant grid in log($\gamma$). Limits of the integration ($\gamma_{\min}$, $\gamma_{\max}$) and grid points per unit in log($\gamma$) are user-defined. For some electron distributions, such as the thermal distribution, setting integration limits may be less intuitive. However, it is necessary within our formalism and implies truncated electron distributions at the limits. Computations involving power-law distributions adapt the integration limits from the distribution parameters ($\gamma_{\min}$, $\gamma_{\max}$) and the ones with exponential cutoff use the integration limits ($\gamma_{\min}$, $10\times\gamma_{\max}$) instead. This only applies to the integration limits and does not necessarily reflect the normalisation of electron distributions, which are discussed in Appendix~\ref{sec:ap_edist}.

For tangled magnetic fields, we approximate the angle-averaged computation with $\phi = \arcsin(\pi/4)$. This is the preferred approach in some studies such as \cite{Dodds-Eden10} for a faster computation by avoiding integration over the solid angle.

\subsection{Inverse Compton scattering}

Adapting the formalism of \cite{BlumenthalGould70}, we define a Compton kernel ($F_C$),
\begin{equation}
    F_C(\epsilon, \gamma, \epsilon') = \frac{3\sigma_T c}{4\gamma^2 \epsilon} \left[2q\ln q\!+\!(1\!+\!2q)(1\!-\!q) \!+\! \frac{(\Gamma_e q)^2(1\!-\!q)}{2(1\!+\!\Gamma_e q)} \right],
\end{equation}
where $\Gamma_e = 4\epsilon\gamma/m_e c^2$ and $q=\epsilon'/\Gamma_e(\gamma m_e c^2 - \epsilon')$, $\sigma_{\textrm{T}}$ is the Thomson scattering cross-section. Here, $\epsilon$ and $\epsilon'$ respectively refer to the incident and the scattering photon energies, that is, $\epsilon'=h\nu$ with $h$ as the Planck constant. We then calculate the Compton emissivity ($j_{\nu,\tic}$) with a double integral along $\epsilon$ and $\gamma$,
\begin{equation}
    j_{\nu,\tic} = \frac{h \nu}{4\pi} \int_{\epsilon} h n_p(\epsilon/h) \int_{\gamma} n_e(\gamma) F_C(\epsilon, \gamma, \epsilon') d\gamma d\epsilon,
\end{equation}
where $n_p$ is the seed photon density.

The corresponding generic low-level utility function operates on a given arbitrary grid of seed photon density and electron distribution. While this is the most computationally expensive task in our code, it can be massively parallelised. Therefore, we also provide an OpenCL implementation using the \textsc{pyopencl} package\footnote{\url{https://github.com/inducer/pyopencl}}, which executes the computations on a GPU and is hardware agnostic.

\subsection{Ray-tracing}

To simplify the radiative transfer calculations, either the slab approximation or a homogeneous sphere geometry is used for the analytical modelling of astrophysical synchrotron sources. Without these assumptions, radiative transfer calculations either become very complex or impossible. Ray-tracing is an alternative method for numerical calculations, for which imaginary rays passing through a source and exiting towards the line of sight are drawn. Specific intensities are calculated along these rays numerically. The specific intensities at the source surface are then used to estimate its brightness.

Our code provides a basic support for ray tracing to enable the modelling of synchrotron sources with complex geometries. The corresponding utility function is capable of parallel computation of multiple rays with the same length and fixed grid resolution, $\Delta x$. A single ray tracing has the form 
\begin{equation}
    I_{\nu(k)} = \sum^{k}_{i=0}  \Delta x (j_{\nu_{(i)}}/\alpha_{\nu_{(i)}}) \left( \prod^{k}_{i} \left(1-\exp(-\alpha_{\nu_{(i)}}\Delta x) \right) \right), \label{eq:rtrace}
\end{equation}
where $I_{\nu(k)}$ is the specific intensity at grid point $k$ and frequency $\nu$, and the boundary condition is set by $I_{\nu(0)}=0$. As such, this function stores the specific intensities along the rays although the stored data are not necessary to calculate synchrotron luminosity. Specific intensities on a grid can then be used to compute seed photon densities for SSC emission as described in Sect.~\ref{sec:radialsp}.

\subsection{Geometries}

We provide two alternative geometries with our code. These demonstrate how to construct source models with the built-in utility functions.

\subsubsection{Homogeneous sphere} \label{sec:homosp}

A homogeneous sphere is the simplest assumption one can make on the geometry of compact synchrotron sources. It is therefore one of the most commonly used. As an advantage to analytical modelling, our implementation can compute synchrotron SEDs from a wide selection of electron distributions. Radiative transfer on a homogeneous sphere can be written as,
\begin{equation}
    L_{\nu} = 4\pi \int^R_0 \frac{2\pi r j_\nu}{\alpha_\nu} \left[1-\exp{\left(-2\alpha_{\nu}\sqrt{R^2-r^2}\right)}\right] \, dr,
\end{equation}
where $R$ is the radius of the sphere\footnote{We always refer to the observed $L_{\nu}$ unless otherwise indicated.}. We use the analytical form of this integral because its computation is faster. The luminosity integral reduces to
\begin{equation} \label{eq:thin_sync}
    L_{\nu\textrm{,thin}} = \frac{4\pi}{3} R^3 (4\pi j_{\nu})
\end{equation} 
and to 
\begin{equation} \label{eq:thick_sync}
    L_{\nu\textrm{,thick}} = 4\pi \left(\frac{\pi j_{\nu}R^2}{\alpha_{\nu}}\right)
\end{equation} 
when the emission is optically thin and optically thick, respectively.

In this model, seed photon densities for SSC calculations are estimated using the equation 
\begin{equation} \label{eq:seedp}
    n_p(\nu)=\frac{L_{\nu}}{4\pi c h^2 \nu R^2}. 
\end{equation}
We assume the medium is transparent to the IC emission. Hence, the IC luminosity ($L_{\nu,\tic}$) is equivalent to Eq.~(\ref{eq:thin_sync}) replacing $j_{\nu}$ with $j_{\nu,\tic}$.

We include a Doppler beaming formalism shared by both geometrical models. If we define a Doppler factor, 
\begin{equation}
    \delta \equiv \Gamma^{-1}/(1-\beta \sin\Phi)
\end{equation}    
with $\Gamma^{-1}=\sqrt{1-\beta^2}$, for a source moving with velocity $\beta(=v/c)$ at an angle $\Phi$ to the observer, the implementation is as follows,
\begin{align}
    L_{\tsyn}(\nu) &= \delta^n L_{\tsyn}' (\nu'), \\
    L_{\tssc}(\nu) &= \delta^n L_{\tssc}' (\nu', n_p'), \\
    L_{\tic}(\nu) &= \delta^n L_{\tic}' (\nu', n_p^\star),
\end{align}
where $\nu' = \nu / \delta$ and $n$ is a geometry-dependent parameter; typically $n=3$ for a point source. The quantities $L'$, $\nu'$ , and $n_p'$ refer to the values in the rest frame of the source. The photon density of an external source ($n_p^\star$) needs to be supplied by the user because it is dependent on the modelling geometry. In our formulation, this quantity is in the rest frame and at the location of the main source. This is implemented in the form of a function decorator in Python, and therefore the definition can be adapted to different scenarios; for example, extra-galactic sources with a redshift.

\subsubsection{Radial sphere} \label{sec:radialsp}

The treatment of inhomogeneous spherical sources is discussed in \cite{Gould79} and \cite{Marscher77}, and a full theoretical framework is introduced in \cite{BandGrindlay85} for power-law spheres, in which electron density and magnetic field radially scale in the form of a power law. Our implementation of a radial sphere is similar to those concepts but generalised to a spherical source with any given electron distribution, and it is possible to define an arbitrary radial scaling of electron density and magnetic field.  

To estimate synchrotron emission from the sphere, we first calculate $j_\nu(r)$ and $a_\nu(r)$ on a one-dimensional regular grid where the number of grid points is specified by the user $n_r=R/\Delta r$. This is only valid for $R_{\tin}<r<R$ where $R_{\tin}$ and $R$ are respectively the inner and the outer radii of the source. $j_\nu(r)$ and $a_\nu(r)$ are set to zero beyond these limits. We then interpolate $j_\nu$ and $a_\nu$ to a regular Cartesian grid of size $(R, 2R)$ and of resolution $\Delta r$. In the end, we perform ray tracing with the utility function provided in our code; see Eq.~(\ref{eq:rtrace}). From here on, computing the synchrotron luminosity is straightforward with 
\begin{equation}
    L_\nu = 4\pi \sum I_\nu(r) 2\pi r \Delta r.
\end{equation}

Unlike the formalism used in the homogeneous sphere model  (Eq.~(\ref{eq:seedp})), this geometrical model provides proper numerical estimates of the seed photon densities within a spherical source. Even for a homogeneous sphere (physically, not referring to the corresponding model in Sect.~\ref{sec:homosp}), the  formalism is not exact and seed photon densities additionally vary with radius. Here, we describe our methodology while the numerical result is explored in Sect.~\ref{sec:homo_SSC}. 

Our methodology is similar to the concept of \cite{BandGrindlay85}, in which they apply a ray-tracing solution for rays passing through a single point in radius from all directions. The specific intensities of each ray at that point are then normalised. In our case, we only apply ray tracing along a single direction as described above, and then we normalise specific intensities at grid points falling into a layer of finite size in radius. 

Mathematically, this computation can be summarised as
\begin{equation}
    n_p(\nu,r) = \left( \frac{2\pi}{h^2\nu c} \right) \frac{\sum^{\mathcal{R}(r)} I_{\nu_{(i,j)}} \sin\theta_{(i,j)}}{\sum^{\mathcal{R}(r)} \sin\theta_{(i,j)}},
\end{equation}
where $\mathcal{R}(r)$ represents the grid points in the range $r-\Delta r/2 < r_{(i,j)} < r+\Delta r/2$. Finally, we sum the contributions of each layer to the total SSC emission using the computed $n_p(\nu,r)$. \footnote{This solution is only valid for an isotropic synchrotron emissivity.}

\subsection{Temporal evolution} \label{sec:temp}

We include a basic temporal model for high-energy electrons used in \cite{Dodds-Eden10} to investigate the temporal evolution of a bright and complex multi-wavelength flare of \sgr. With this formulation, it is possible to evolve and calculate custom electron distributions and/or to build time-dependent models simulating adiabatic/synchrotron cooling with particle injection, by writing  
\begin{equation} \label{eq:cool}
    \frac{\partial N_e(\gamma, t)}{\partial t} = c_{\tinj}(t) N^\star_e(\gamma)  - \frac{N_e(\gamma, t)}{\tau_{\tesc}} + \frac{\partial \Dot{\gamma} N_e(\gamma, t)}{\partial \gamma}.
\end{equation}
The first term in the equation represents the particle injection into the system with normalised electron distribution $N^\star_e(\gamma)$ and of rate $c_{\tinj}(t)$. The particle escape is given in the second term with a timescale of $\tau_{\tesc}$. The third term simulates the cooling of electrons in the system, where $\Dot{\gamma}$ is the sum of contributions from adiabatic and synchrotron cooling. Instantaneous adiabatic losses when geometry is reduced to a sphere is given by
\begin{equation} \label{eq:adiab_cool}
    -\Dot{\gamma}_\text{adiab} = \gamma \frac{\mathrm{d} \ln(R(t))}{\mathrm{d}t} = \frac{\gamma c \beta_{\exp} }{R(t)},
\end{equation}
where $\beta_{\exp}$ is the expansion speed of the source in units of $c$, and $R$ is the instantaneous radius of the source at time $t$. Angle-averaged synchrotron losses can be written as
\begin{equation} \label{eq:sync_cool}
    -\Dot{\gamma}_{\tsyn} = \gamma^2 \left(\frac{4\sigma_T}{3 m_e c}\right) \left(\frac{B^2}{8\pi}\right) = \frac{\gamma^2 B^2} {7.74 \times 10^8} \text{ s}^{-1}.
\end{equation}

As the temporal evolution of electrons can be formulated in many different ways, our utility function only performs one step integration from one state to the following state after a time-step of $\Delta t$. Appendix~\ref{sec:num_temporal} details our numerical formulation. We provide Python interfaces for selected scenarios as guidance, such as the examples in Sects.~\ref{sec:ad_cool} and \ref{sec:synccool}.

The effects of IC processes are not included in this model. Therefore, there is a hidden assumption that IC cooling is negligible compared to other cooling processes. With a first-order approximation, the relative contribution of IC to synchrotron cooling can be estimated from the ratio of the seed photon energy density ($u_p$) to the magnetic energy density ($u_B=B^2/8\pi$). This approach can serve as a validity check for the compatibility of a given scenario with this model. 

In its most basic form, IC cooling can be similarly implemented by introducing an effective energy density ($u_\mathrm{eff}=u_p+u_B$) to Eq.~(\ref{eq:sync_cool}) at the cost of an additional computation time associated with the estimate of $u_p$, which needs to be updated at each integration step. However, there are additional concerns associated with it such as radiation pressure and multiple scatterings when IC processes are dominant. Implementing these physics would require much more complex formulations, which are beyond the scope of this study. If they are not implemented, there is no clear cut to when these physics will be problematic. Hence, we choose not to explicitly support IC cooling in a generic method with no restrictions.

\section{Tests and examples} \label{sec:test}

\begin{figure}
    \centering
    \resizebox{\hsize}{!}{\includegraphics{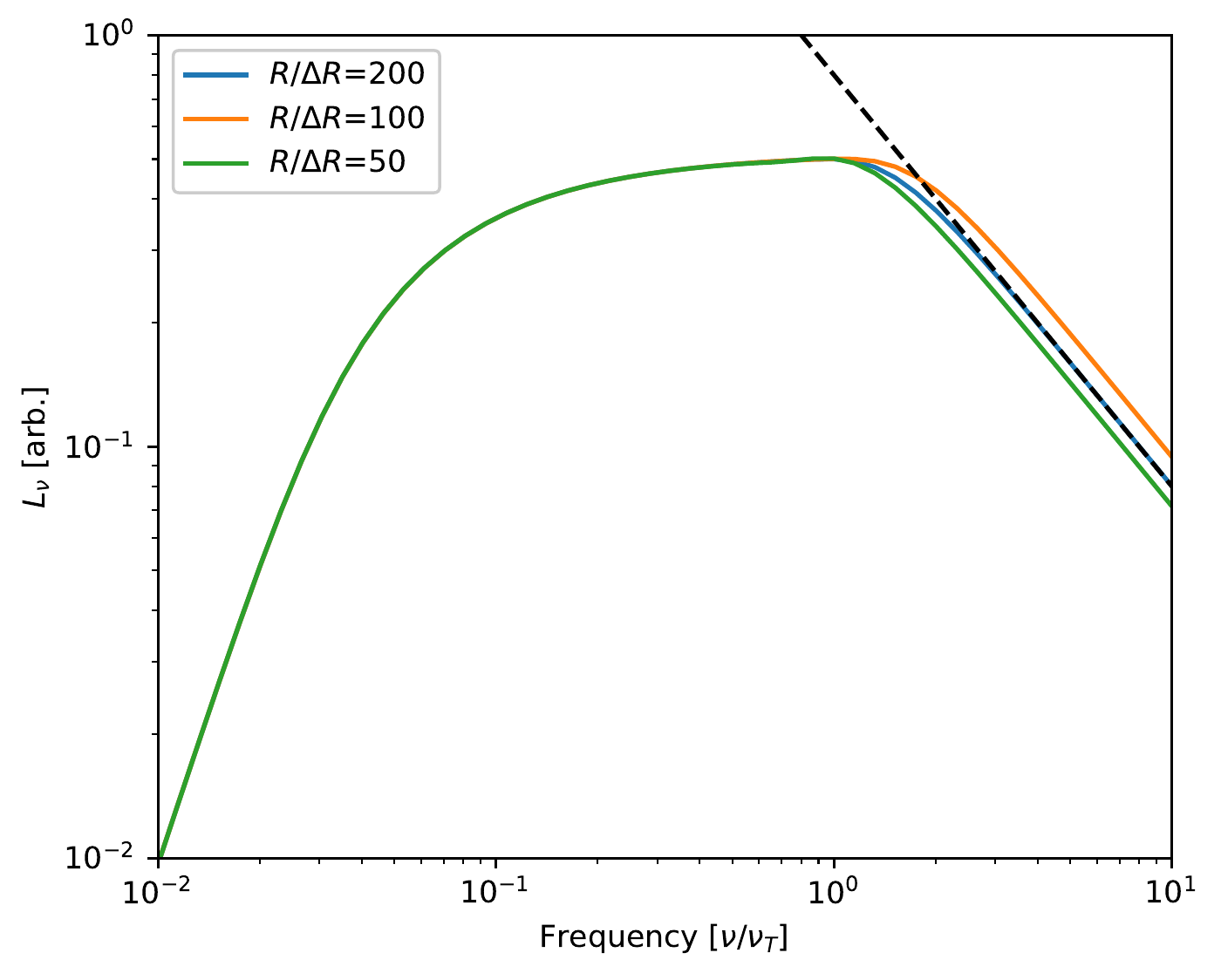}}
    \caption{Synchrotron SED using our radial sphere implementation constructed with parameters $m=1$, $n=2$, $x=50$ and $\alpha=1$ as formulated for a power-law sphere in \cite{BandGrindlay85}. SEDs plotted in different colours refer to calculations at selected radial resolutions ($R/\Delta R$) - 50, 100, and 200 grid points in radius respectively for $x$, $2x,$ and $4x$. The black dashed line is the analytical form of the expected optically thin emission in Eq.~(\ref{eq:BG85_SEDthin}). $\nu_T$ is the numerical estimate of the frequency where the SED peaks.}
    \label{fig:BG85_pl}
\end{figure}

\subsection{Synchrotron emission from a power-law sphere} \label{sec:pSSC}

To demonstrate our radial sphere implementation, we reproduce the solution of \cite{BandGrindlay85} (in their Figure 1b) corresponding to their power-law sphere formulation. Using the definition provided by these latter authors, the magnetic field in the sphere scales with $B = B_0 (r/R_{\tin})^{-m}$ and $m=1$; the electron density scales with $n_e = n_{e,0} (r/R_{\tin})^{-n}$ and $n=2$.  The index of the power-law electron distribution is  $p=3$ ($p=2\alpha+1$). Finally, the radius is parameterised as $x=R/R_{\tin}$ with $x=50$. The choice of $B_0$ and $n_{e,0}$ is not relevant for the solution as long as the effects of ($\gamma_{\min}$, $\gamma_{\max}$) are negligible in a selected frequency range.

The resulting SED shown in Fig.~\ref{fig:BG85_pl} has three distinct regimes from higher to lower frequencies: (1) The sphere is optically thin above $\nu_T$, which we calculate numerically, and is the transition frequency to the optically thin emission. (2) The partially opaque regime can be interpreted as a superposition of SEDs corresponding to different layers along the radius. (3) The sphere is fully opaque below a certain frequency ($\nu \lesssim 3\nu_T\times10^{-2}$) and produces an SED with $L_\nu \propto \nu^{2.5}$.

There is no exact analytical solution for this particular problem\footnote{\cite{Marscher77} provide an analytical formalism to describe the full shape of the SED using a slab approximation. While the accuracy is reasonable for most applications, \cite{BandGrindlay85} note discrepancies with respect to the numerical solution.}. \cite{BandGrindlay85} provide an analytical form describing the optically thin regime as 
\begin{equation} \label{eq:BG85_SEDthin}
    L_\nu=\frac{4\pi R_{\tin}^3}{3}\left(\frac{x^{3-s} - 1}{3-s}\right) [4\pi j_\nu (R_{\tin})],
\end{equation} 
where $s = n+m(a+1)$. This is the dashed line in Fig.~\ref{fig:BG85_pl}. 

One reason to present this particular example is to further discuss our ray-tracing solution. We provide our solution computed at different radial resolutions with multiples of $x$ in the figure. While all three computations agree below $\nu_T$, the reproducibility of the optically thin SED requires  common sense as to the choice of radial resolution, because it is necessary to resolve the innermost region near $R_{\tin}$ as shown in Eq.~(\ref{eq:BG85_SEDthin}). The difference can be more dramatic with steeper slopes either in the magnetic field or in the electron density along the radius and a finer resolution may be required.

\begin{figure}[!ht]
    \centering
    \resizebox{\hsize}{!}{\includegraphics{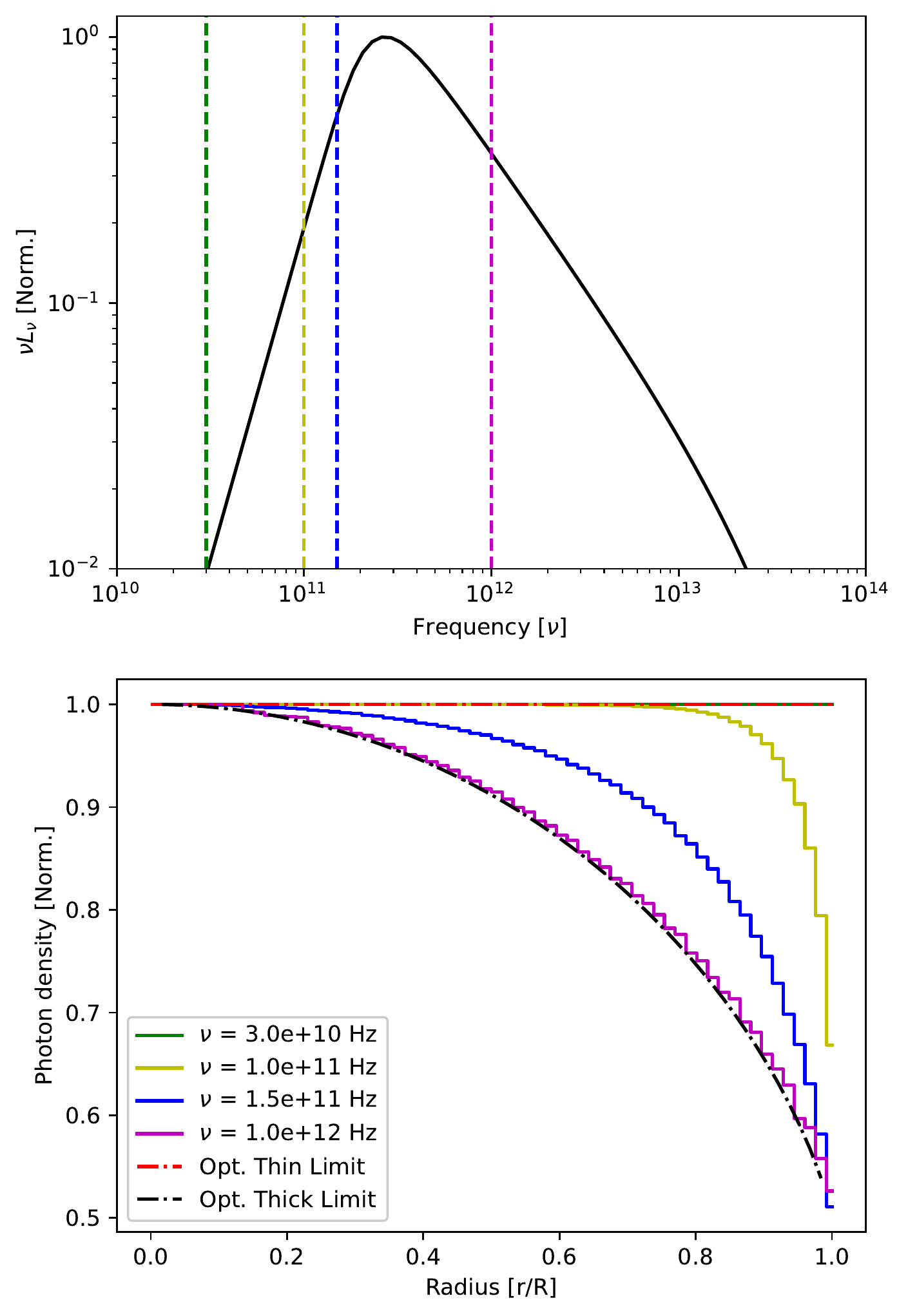}}
    \caption{Top: Representative synchrotron SED where vertical dashed lines are equivalent to the colour-coded frequencies in the bottom plot. Bottom: Numerical solution to the photon densities in a homogeneous sphere at the selected frequencies. The optically thin and thick limits are shown with the two dot-dashed lines.}
    \label{fig:PhHomoSp}
\end{figure}

\subsection{Synchrotron self-Compton emission from a homogeneous sphere} \label{sec:homo_SSC}

One of the challenging steps for accurate SSC calculations is the estimation of seed photon densities (from synchrotron emission) within a given source. A common approach is to use Eq.~(\ref{eq:seedp}) as in our homogeneous sphere model for simplicity, but this is only valid at a radius beyond the surface of the source. Proper estimation of seed photon densities at a given location (\textbf{r}) requires summation of contributions from different locations (\textbf{r'}) in an entire source. Hence, an exact analytical solution is not possible and numerical calculations are computationally expensive. 

Briefly, we only discuss our results for a homogeneous sphere complementary to the description in Sect.~\ref{sec:radialsp}. There are two limits to the photon densities: 

\begin{itemize}
    \item At frequencies where the sphere is opaque, the mean free path of photons is small compared to the size of the source. Hence, photon densities in this limit are constant except near the outer regions of the sphere. 
    \item At frequencies where the sphere is optically thin, the radial profile is given by
    \begin{equation}
       \Phi(x) = \frac{2}{3} \left(\frac{1-x^2}{2x} \ln \left|\frac{1+x}{1-x} \right| +1 \right),
    \end{equation}
where $x$ is the normalised radius $r/R$ \citep{BandGrindlay85}. 
\end{itemize}

\begin{figure}
    \centering
    \resizebox{\hsize}{!}{\includegraphics{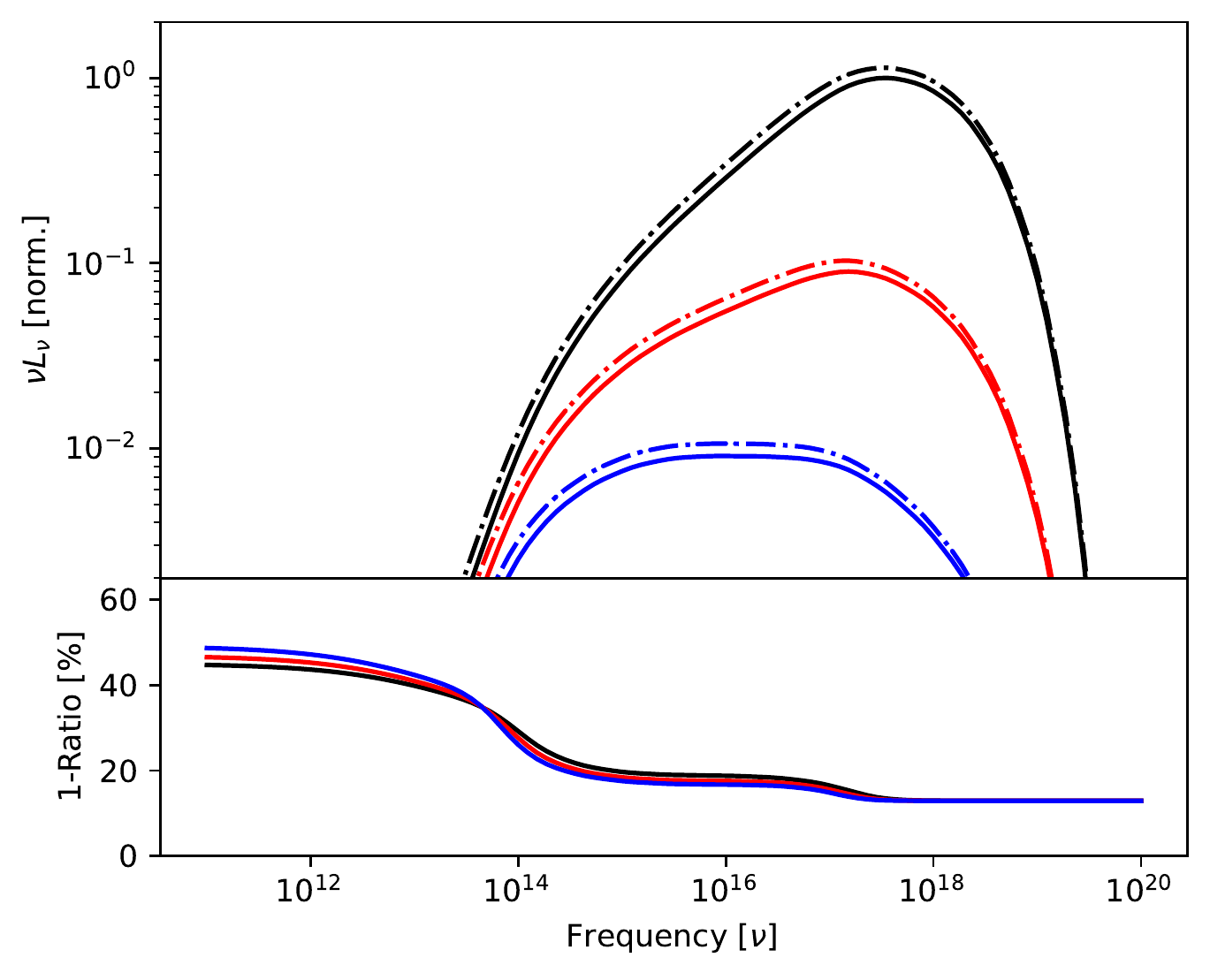}}
    \caption{Synchrotron self-Compton SED for a homogeneous sphere with $p$ values 2, 2.5, and 3 shown in different colours. Solid lines represent the SEDs with the traditional photon density estimates and dashed lines those with the numerical photon densities as described in the text. The bottom plot shows the difference between the two methods.}
    \label{fig:SSCComp}
\end{figure}

We show our results for selected frequencies in Fig.~\ref{fig:PhHomoSp}. Our calculations compare well with the expected radial profiles in both limits. This result alone may not be very intuitive. Therefore, we also compute the SSC spectrum for $p$ of 2, 2.5, and 3 in Fig.~\ref{fig:SSCComp}. Compared to the approximation with Eq.~(\ref{eq:seedp}), our code predicts 15\%-20\% excess SSC emission in the optically thin regime and about 50\% excess in the optically thick regime with small variations at different $p$ values. The difference can be incorporated in $n_e$ and $\gamma_{\min}$ in this specific case. As a result, for a homogeneous sphere, this calculation is unnecessarily expensive: it roughly scales linearly with the radial resolution.

\subsection{Adiabatic cooling} \label{sec:ad_cool}

The adiabatic cooling model introduced in \cite{vdL66} is still one of the most commonly used models to describe the temporal evolution of compact synchrotron sources at longer wavelengths (typically from radio to sub-mm) due to its simplicity. In essence, the model is a recipe to produce synchrotron flares without injection of electrons to the system. As the source expands, the self-absorbed peak of the synchrotron emission propagates towards lower frequencies with the combination of decreasing electron density ($\propto 1/R^3$) and magnetic field ($\propto 1/R^2$). This evolution describes the rise of the flare. As the self-absorbed peak propagates through a given frequency, flux starts to drop due to adiabatic cooling, and therefore the decay of the flare sets in. Naturally, this formalism predicts a certain time-delay from shorter to longer wavelengths.

Aside from its simplicity and fast numerical evaluation, there are some inherent drawbacks to the model. The model can only describe the temporal evolution of the synchrotron SED arising from a plain power-law electron distribution and it is not possible to directly infer the physical properties of the source. In addition, physical implications of the magnetic field, such as  synchrotron cooling, are not taken into account, whereas in our numerical implementation, the temporal model directly operates on the electron population itself and can simulate the adiabatic and synchrotron cooling simultaneously.

\begin{figure}
    \centering
    \resizebox{\hsize}{!}{\includegraphics{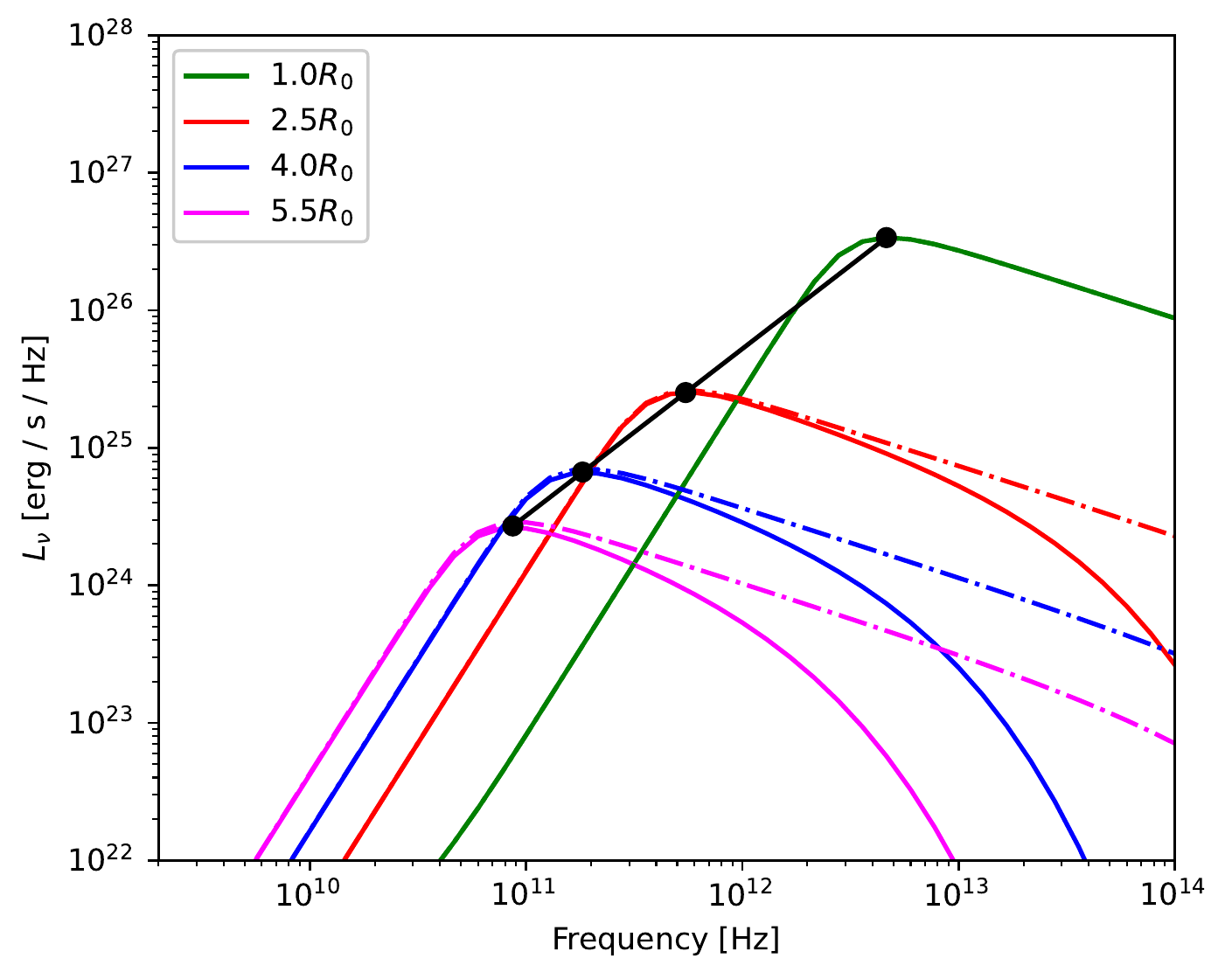}}
    \caption{Temporal evolution of synchrotron SEDs calculated at selected radii with the utility function discussed in Sect.~\ref{sec:temp}. We refer to the text for further details. Coloured dashed lines refer to an evolution in which synchrotron cooling is assumed to be negligible, and this would be equivalent to the traditional \cite{vdL66} formalism. Solid lines represent the case in which synchrotron cooling is enabled. The black line is the expected evolution of the self-absorbed peak according to the \cite{vdL66} model.}
    \label{fig:vdL}
\end{figure}

We set up a test problem using the parameters from \cite{YusefZadeh08} as a reference ($p=2$, $B_0=10$ G, $\beta_{\exp}=0.02c$, $R_0=2R_{\tS}$ of \sgr with $R_{\tS}=1.3\times10^{12}$ cm). We set $\gamma_{\max}$ of the initial electron distribution to a large number so it is not a factor in our calculations. We integrate over Eq.~(\ref{eq:cool}) with a time resolution of one second. It is not necessary to develop the full model here. We rather use the scaling relations from \cite{vdL66},
\begin{align}
    L_{\nu,m}(R) &= L_{\nu,m}(R_0) (R/R_0)^{-(4p+6)/(p+4)}, \\
    \nu_m(R) &= \nu_m(R_0) (R/R_0)^{-(7p+3)/(p+4)}, 
\end{align}
where $L_{\nu,m}(R)$ and $\nu_m(R)$ represent the self-absorbed peak of SED at a given radius. In Fig.~\ref{fig:vdL}, we plot results of our numerical integration at selected radii with synchrotron cooling enabled and disabled. We also overplot the evolution of the self-absorbed peak predicted from the \cite{vdL66} model. As seen in the figure, synchrotron cooling introduces a spectral cutoff at high energies which propagates to lower frequencies as the source expands. This is despite the choice of large $\gamma_{\max}$, which has no effect here.

An analytical approximation for the evolution of the high-energy cutoff is provided in Appendix~\ref{sec:app_ad_cool}. The final result is sensitive to the initial magnetic field and implies an upper limit to the magnetic field where the \cite{vdL66} assumption is still valid. In other words, the underlying magnetic field needs to be large enough to produce the observed flux arising from synchrotron processes, but at the same time it needs to be small enough so that the impact of the synchrotron cooling to the system is negligible at the frequency range of interest. This aspect of the model limits its applicability to lower frequencies, typically radio to sub-mm range, where the effects of synchrotron cooling are less prominent.

\subsection{Synchrotron cooling} \label{sec:synccool}

There are two traditional textbook solutions to the synchrotron cooling: (1) A propagating break ($\gamma_{b}$) in the electron distribution during sustained particle injection, and (2) a propagating cutoff ($\gamma_{\max}$) during limited or no injection \citep{Longair11}. In both cases, the formulation is similar (either for the break or the cutoff in the electron distribution respectively) with 
\begin{equation}
    \gamma_{b,\max}\sim1.3\times10^5 (B/10\text{ G})^{-2} (t_{\tsyn}/1\mathrm{ min})^{-1},
\end{equation}
where $t_{\tsyn}$ is comparable to the injection timescale ($t_{\tinj}$, or other dynamic timescale in the system) for the former case and the time after the particle injection for the latter. 

We proceed with a formulation somewhere between the two extremes to demonstrate both cases at once. We consider particle injection in the form of a power-law distribution with $p=2$, $\gamma_{\min}=50$ and $\gamma_{\max}=4\times10^5$. The particle injection rate has a Gaussian profile with $\sigma=5$ min, peaking at $3\sigma$. In this setup, we can assume an effective injection timescale of $t_{\tinj}\sim2.35\sigma/2$. We fix the magnetic field to 20 G.

\begin{figure}
    \centering
    \resizebox{\hsize}{!}{\includegraphics{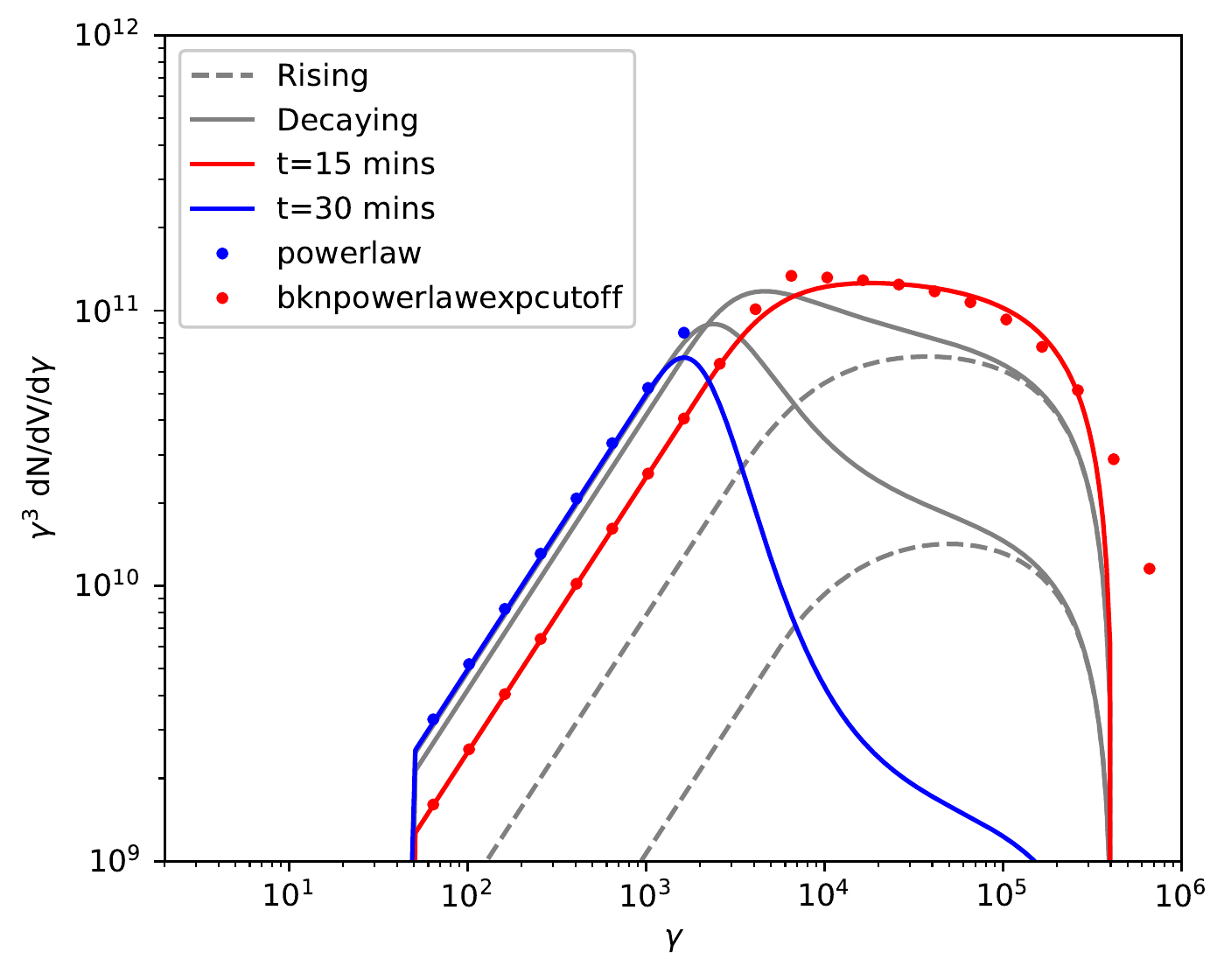}}
    \caption{Evolution of a power-law electron distribution under synchrotron cooling and Gaussian particle injection with parameters noted in the text. Time evolution of the electron distribution is presented as snapshots taken every 5 minutes. Grey dashed lines refer to the rising part of the evolution and solid lines to the decaying part. The red line represents the spectrum at the peak of injection profile ($t=15$ min) whereas red dots are the approximation with a broken power-law distribution. The blue line is the electron distribution 15 minutes after injection peak ($t=30$ min) and blue dots are the approximation with a simple power law.}
    \label{fig:SyncCool}
\end{figure}

We plot the evolution of the electron distribution with this experimental setup in Fig.~\ref{fig:SyncCool}. While it is difficult to describe each time stamp with a standard electron distribution, we pick two cases where an approximation is possible. These refer to the time stamps; injection peak at 15 minutes and when synchrotron cooling `erases' the history of the particle injection at $t\sim30$ minutes.

The former case can be approximated with a broken power-law electron distribution. The break in the electron distribution is estimated from $t_{\tinj}$ and $\gamma_{b}$ using the arguments described above. Additionally, we use an exponential cutoff in the electron distribution setting $\gamma_{\max}=(\gamma_{\max})_{\tinj}/1.5$. The latter case can be approximated with a simple power-law distribution. In this case, $\gamma_{\max}$ is chosen so that there is synchrotron cooling with an effective duration of 15 minutes, that is, the 15 minutes following the peak injection.

As demonstrated here, analytical approximations are possible for this particular setup by reducing the problem to selected time episodes. Similar arguments have been used either to describe flare SEDs linking certain source properties to spectral features \citep{Ponti17,2021arXiv210701096A} or to extract physical properties of sources at certain time intervals \citep{Dallilar17}. However, there are also studies providing evidence for multiple injection and cooling episodes or perturbations in physical properties during flaring episodes. For example, \cite{Dodds-Eden10} investigated a bright \sgr flare with a rapid substructure and linked changes in the magnetic field to the substructure in the multi-wavelength light curves. Numerical tools such as those described here provide the potential to investigate the dynamical evolution of such events in more detail, which is useful for disentangling possible emission scenarios and understanding the history of particle acceleration and/or cooling.

\section{Conclusion}

In this paper, we present \flaremodel, an open-source Python package for one-zone modelling of astrophysical synchrotron sources. We discuss our methodology and the structure of our code along with selected examples to demonstrate the capabilities and options for future modelling efforts.

 Instead of focusing on selected source models, we provide a set of generic low-level utility functions from which users can build their own models. This is a step towards modularity and ease of development in future modelling efforts while still keeping the duration of numerical calculations reasonably low with multi-threading support. As such, basic models can still be executed on low-end hardware but complex models designed for high-end hardware can take advantage of built-in multi-threading support. This approach distinguishes \flaremodel from similar alternatives.

We do not provide detailed benchmarks with this paper but only order-of-magnitude estimates for built-in geometrical models. We use the default grid settings. SEDs in each case are computed for 100 points in frequency on a workstation with an i7-8700 CPU using a single thread. For the homogeneous sphere model, SEDs are computed within 1 ms and 100 ms for synchrotron and SSC, respectively. For the radial sphere model, computations are performed in roughly 100 ms and several seconds for synchrotron and SSC, respectively. When SSC calculations are offloaded to the integrated graphics of this CPU, the execution time reduces to about 1 s.

\begin{acknowledgements}
SvF, \& FW acknowledge support by the Max Planck International Research School. 
GP acknowledges funding from the European Research Council (ERC) under the European Union’s Horizon 2020 research and innovation programme (grant agreement No 865637). 
\end{acknowledgements}


\bibliographystyle{aa}
\bibliography{flaremodel.bbl}

\newcommand{\tth}{\mathrm{th}}
\newcommand{\tpl}{\mathrm{pl}}
\newcommand{\tbpl}{\mathrm{bpl}}

\begin{appendix}

\section{Analytical formulation of implemented electron distributions} \label{sec:ap_edist}

As described in the text, we provide a handful of useful electron distributions with our code and their analytical formulation is given here. The normalised \textit{thermal} electron distribution is implemented as
\begin{equation}
    \frac{\mathrm{d} n_{\tth}}{\mathrm{d}\gamma} = \frac{\gamma (\gamma^2-1)^{1/2}}{\Theta_e K_2(1/\Theta_e)} \exp\left(-\frac{\gamma}{\Theta_e}\right),
\end{equation}
where $\Theta_e(=k_B T/m_e c)$ is the dimensionless electron temperature and $K_2$ is the modified Bessel function of the second kind.

For non-thermal electron distributions, there are two main options. For a {power-law} electron distribution, we use the form
\begin{equation}
    \frac{\mathrm{d} n_{\tpl}}{\mathrm{d}\gamma}= N_{\tpl} \gamma^{-p},  \qquad \text{for } \gamma_{\min} \leq \gamma \leq \gamma_{\max},
\end{equation}
where $p$ is the exponent of the distribution, and $\gamma_{\min}$ and $\gamma_{\max}$ are lower and higher energy limits,  respectively. We define the {broken power-law} distribution as the combination of two power-law distributions with exponents $p_1$ and $p_2$,  below and above the break in gamma ($\gamma_b$), respectively. This latter  has the form,
\begin{equation} \label{eq:edist_bpl}
    \frac{\mathrm{d} n_{\text{bpl}}}{\mathrm{d}\gamma}= N_{\tbpl} \times
    \begin{cases}
        \gamma^{-p_1}, & \text{if } \gamma_{\min} \leq \gamma\leq \gamma_b,\\
        \gamma^{-p_2} \gamma_b^{p_2-p_1}, & \text{if } \gamma_{b} < \gamma \leq \gamma_{\max}.
    \end{cases}
\end{equation}
The normalisation factors for these distributions are respectively
\begin{equation} \label{eq:norm_pl}
    N_{\tpl} = \frac{p-1}{\gamma_{\min}^{1-p}-\gamma_{\max}^{1-p}},
\end{equation}
and
\begin{equation} \label{eq:norm_bpl}
    N_{\tbpl} = \left[\left(\frac{\gamma_{\min}^{1-p_1} - \gamma_{b}^{1-p_1}}{p_1-1}\right) + \gamma_b^{p_2-p_1}\left(\frac{\gamma_{b}^{1-p_2} - \gamma_{\max}^{1-p_2}}{p_2-1}\right)\right]^{-1}.
\end{equation}
In addition, we provide an alternative form for both power-law electron distributions with an \textit{exponential cutoff} at $\gamma_{\max}$. Explicitly, this is implemented as
\begin{equation}
    \frac{\mathrm{d} n_{\text{expc}}}{\mathrm{d}\gamma} = 
        \left(\frac{dn}{d\gamma}\right)_{\tpl,\tbpl} \exp\left(-\frac{\gamma}{\gamma_{\max}}\right),  \qquad \text{for } \gamma_{\min} \leq \gamma.
\end{equation}
\noindent In this case, we still use the corresponding normalisation factors of each distribution given in Eqs.~(\ref{eq:norm_pl}) and (\ref{eq:norm_bpl}). Although this formulation removes the necessity of an upper limit on $\gamma$, the synchrotron routines enforce an upper limit to integration at $10\times\gamma_{\max}$ for practicality. 

Lastly, we include the \textit{kappa} electron distribution as described in \cite{Pandya16}. However, the normalisation of this electron distribution is not exact in their formulation. The error in selected cases is provided in their paper. We do not go into further details here.

\section{Synchrotron emissivity and absorption coefficients}

\begin{figure}
    \centering
    \resizebox{\hsize}{!}{\includegraphics{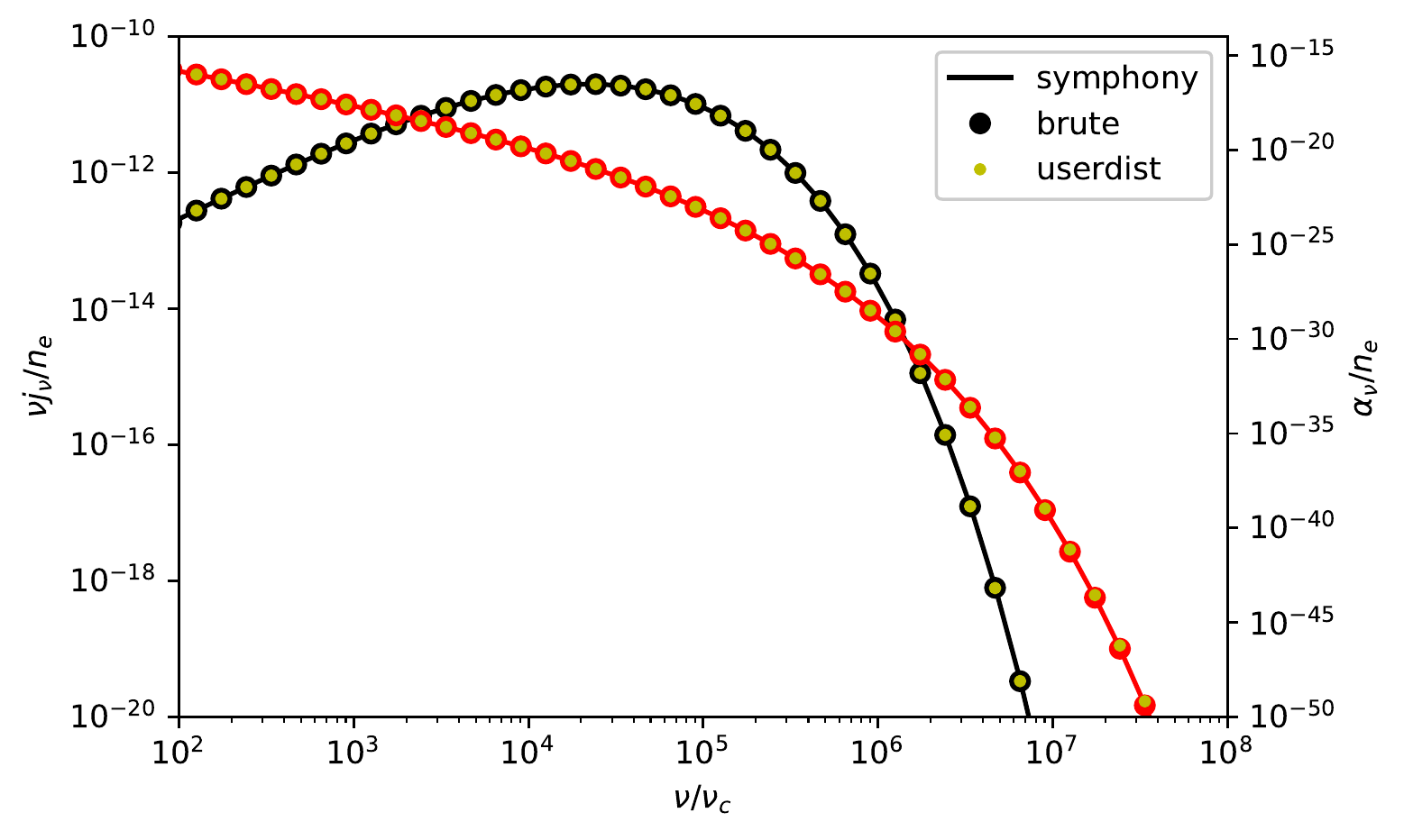}}
    \resizebox{\hsize}{!}{\includegraphics{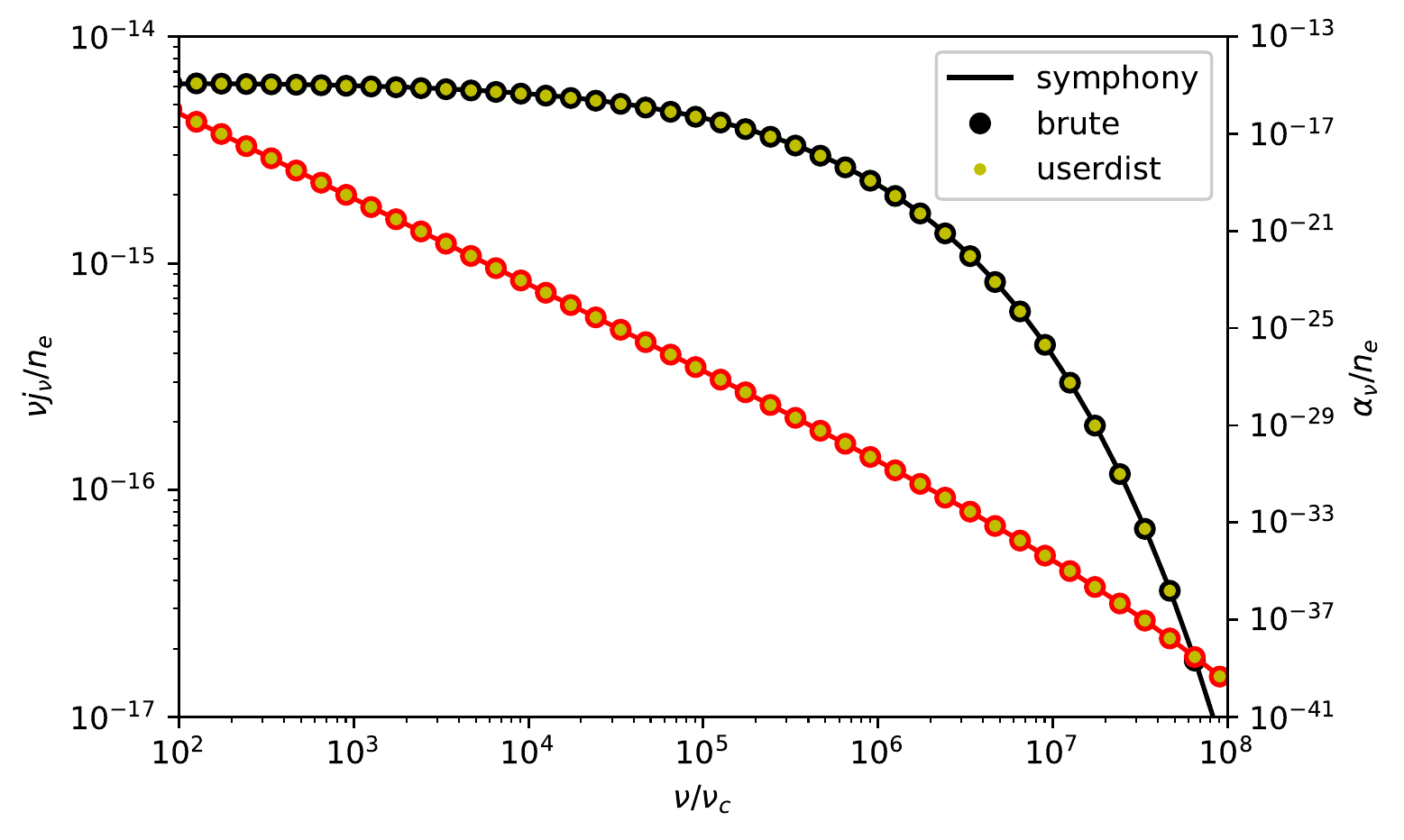}}
    \caption{Numerical computation of synchrotron emissivity (black) and absorption coefficient (red) with the \textit{brute} and \textit{userdist} methods in our code. Thermal (top) and power-law (bottom) electron distributions are shown separately with the parameters specified in the text. The results with \textsc{symphony} are overplotted.}
    \label{fig:SynJA}
\end{figure}

 In Fig.~\ref{fig:SynJA}, we compare our numerical computations of synchrotron emissivity and absorption coefficients with the publicly available \textsc{symphony} \citep{Pandya16} code for thermal ($\Theta_E=30$) and power-law ($p=3$, $\gamma_{\min}=1$, $\gamma_{\max}=10^3$) distributions with $B=10$ G and $\phi=\pi/4$. Numerical integration over the electron distributions with discontinuities is problematic and is also the case with \textsc{symphony}. We can only avoid this complication by setting proper integration ranges with our \textit{brute} method. 
 
 Hence, we use the convenient exponential cutoff option at $\gamma_{\max}$ in \textsc{symphony} and equivalent implementation in our code. This also motivates setting $\gamma_{\min}(\sim\sqrt{\nu/\nu_c})$ low enough so that it is not a factor in this comparison. Our computations are well within a few percent in the range of the plots with the exception of the \textit{userdist} method and then only for $\nu/\nu_c < 10^6$. This originates from the estimate of $\mathrm{d} n(\gamma)/\mathrm{d}\gamma$ in Eq.~(\ref{eq:abs_coef}). This calculation is always more accurate using an analytical form of the distribution. Therefore, we promote the \textit{brute} method as the default while the \textit{userdist} method remains as an option for the electron distributions with no analytical form.

\section{Synchrotron cooling in the presence of adiabatic expansion} \label{sec:app_ad_cool}

When Eq.~(\ref{eq:sync_cool}) is integrated under the assumption of a spherical expansion at a constant speed, i.e. $R(t)=R_0+\beta c$t, the result is
\begin{equation}
    \gamma_{\max}(t) = \frac{3\times7.74\times10^{8}}{{B_0}^2 t_{\exp}} \left(1-\left(\frac{R_0}{R(t)}\right)^3\right)^{-1} \sim \frac{2.3\times10^{9}}{{B_0}^2 t_{\exp}},
\end{equation}
where $t_{\exp} = R_0/\beta c$ and $\gamma_{\max}$ approaches its asymptotic value for $1<<R_0/R$. For the choice of parameters in Sect.~\ref{sec:ad_cool}, $t_{\exp}$ is about $4.2\times10^3$ seconds. With the critical frequency defined in Eq.~(\ref{eq:nu_crit}), we can estimate the time evolution of $\nu_{\max}$ as
\begin{equation}
    \nu_{\max}(t) \sim 0.3 \times \nu_c(\gamma_{\max}, B, \pi/4) \sim 3\times10^{14} \left(\frac{B_0}{10 \text{ G}}\right)^{-3}\left(\frac{R_0}{R(t)}\right)^2 \text{ Hz.}
\end{equation}

There are two main points to emphasise as a result of this formulation. Synchrotron cooling is effective only during the early stages of the expansion with an effective duration on the order of $t_{\exp}$. This determines the high energy cutoff in the electron distribution. At the same time, magnetic flux conservation implies a rapidly decaying magnetic field which pushes $\nu_{\max}$ originating from synchrotron cooling to even lower frequencies as the source expands. 

\section{Numerical scheme used in the temporal evolution} \label{sec:num_temporal}

The temporal evolution model introduced in Sect.~\ref{sec:temp} is a two-dimensional problem. We solve the equation in two steps. Firstly, we estimate the quantities on the right-hand side of the equation in $\gamma$ grid at time $t$. We then calculate electron distribution at $t+\Delta t$ with the forward Euler method.

Going in the reverse order, the time integration is discretised in the form,
\begin{equation}
N^+_{e\mathrm{,i}} = N^-_{e\mathrm{,i}} + (A^-_\mathrm{i} + B^-_\mathrm{i})\Delta t,
\end{equation}
where i denotes the grid points in $\gamma$. We use the notation of minus ($-$) and plus ($+$) respectively for quantities at $t$ and $t+\Delta t$. The quantities $A^-_{\mathrm{i}}$ and $B^-_{\mathrm{i}}$ are separately written in an open form below,
\begin{align}
    A^-_\mathrm{i} &= c_\tinj {N^\star_{e\mathrm{,i}}} - \frac{N_{e\mathrm{,i}}}{\tau_\tesc},\\
    B^-_\mathrm{i} &= \frac{\Dot{\gamma}_\mathrm{i+1}N_{e\mathrm{,i+1}}-\Dot{\gamma}_\mathrm{i}N_{e\mathrm{,i}}}{\gamma_\mathrm{i+1} - \gamma_\mathrm{i}}. \label{eq:cool_-}
\end{align}

Here, we drop the plus and minus notation for readability, because all the quantities are at time $t$. The term $A^-_{\mathrm{i}}$ is reserved for particle injection and escape, and $B^-_{\mathrm{i}}$ for adiabatic and synchrotron cooling. Calculating $\Dot{\gamma}$ is straightforward with the sum of Eqs.~(\ref{eq:adiab_cool}) and (\ref{eq:sync_cool}). 

In the end, with the assumption of a spherical source, the corresponding utility function updates the radius ($R$) and the magnetic field ($B$) in-place with the equations,
\begin{align}
    R^+ &= R^- + c \beta_{\exp} \Delta t, \\
    B^+ &= B^- (R^-/R^+)^2.
\end{align}

As a word of caution, $\Delta t$ is the time-step of the integration, not simply any time interval. The utility function provided with our code can only perform one step integration as described in this section. The proper choice of $\Delta t$ and the formulation of the temporal evolution remains at the discretion of the user. 

\end{appendix}

\end{document}